\documentclass[conference]{IEEEtran}
\IEEEoverridecommandlockouts

\usepackage{caption}
\usepackage{xcolor}
\usepackage{colortbl}
\definecolor{rowcolor}{rgb}{0.85, 0.95, 0.9}

\usepackage{amsmath,amssymb,amsfonts}
\usepackage[ruled, vlined]{algorithm2e} 
\usepackage{graphicx}
\usepackage{textcomp}
\usepackage{multirow}
\usepackage{float}
\usepackage{booktabs}
\usepackage{microtype}
\usepackage{multicol}
\usepackage{subcaption}
\usepackage{diagbox}
\usepackage{tcolorbox}
\usepackage{makecell}
\usepackage{tikz}
\usetikzlibrary{trees}
\usepackage{hyperref}
\hypersetup{
  colorlinks   = true, 
  urlcolor     = blue, 
  linkcolor    = black, 
  citecolor    = black 
}

\usepackage{arydshln} 

\def\BibTeX{{\rm B\kern-.05em{\sc i\kern-.025em b}\kern-.08em
    T\kern-.1667em\lower.7ex\hbox{E}\kern-.125emX}}
\newcommand{\ourmethod}{\textsc{SpeechPrune}}
\usepackage{url}
\usepackage[
backend=biber,
style=ieee,
citestyle=numeric-comp,
maxbibnames=5,
maxcitenames=1,
minnames=1,
doi=false,isbn=false,url=false,eprint=false
]{biblatex}

\defbibheading{bibliography}[\refname]{}

\DeclareSourcemap{
	\maps[datatype=bibtex, overwrite=true]{
		\map{
			\step[fieldsource=booktitle,
			match=\regexp{.*Interspeech.*},
			replace={Proc. Interspeech}]
			\step[fieldsource=journal,
			match=\regexp{.*INTERSPEECH.*},
			replace={Proc. Interspeech}]
			\step[fieldsource=booktitle,
			match=\regexp{.*ICASSP.*},
			replace={Proc. ICASSP}]
			\step[fieldsource=booktitle,
			match=\regexp{.*icassp_inpress.*},
			replace={Proc. ICASSP (in press)}]
			\step[fieldsource=booktitle,
			match=\regexp{.*Acoustics,.*Speech.*and.*Signal.*Processing.*},
			replace={Proc. ICASSP}]
			\step[fieldsource=booktitle,
			match=\regexp{.*International.*Conference.*on.*Learning.*Representations.*},
			replace={Proc. ICLR}]
			\step[fieldsource=booktitle,
			match=\regexp{.*International.*Conference.*on.*Computational.*Linguistics.*},
			replace={Proc. COLING}]
			\step[fieldsource=booktitle,
			match=\regexp{.*SIGdial.*Meeting.*on.*Discourse.*and.*Dialogue.*},
			replace={Proc. SIGDIAL}]
			\step[fieldsource=booktitle,
			match=\regexp{.*International.*Conference.*on.*Machine.*Learning.*},
			replace={Proc. ICML}]
			\step[fieldsource=booktitle,
			match=\regexp{.*North.*American.*Chapter.*of.*the.*Association.*for.*Computational.*Linguistics:.*Human.*Language.*Technologies.*},
			replace={Proc. NAACL}]
			\step[fieldsource=booktitle,
			match=\regexp{.*Empirical.*Methods.*in.*Natural.*Language.*Processing.*},
			replace={Proc. EMNLP}]
			\step[fieldsource=booktitle,
			match=\regexp{.*Association.*for.*Computational.*Linguistics.*},
			replace={Proc. ACL}]
			\step[fieldsource=booktitle,
			match=\regexp{.*Automatic.*Speech.*Recognition.*and.*Understanding.*},
			replace={Proc. ASRU}]
			\step[fieldsource=booktitle,
			match=\regexp{.*Spoken.*Language.*Technology.*},
			replace={Proc. SLT}]
			\step[fieldsource=booktitle,
			match=\regexp{.*Speech.*Synthesis.*Workshop.*},
			replace={Proc. SSW}]
			\step[fieldsource=booktitle,
			match=\regexp{.*workshop.*on.*speech.*synthesis.*},
			replace={Proc. SSW}]
			\step[fieldsource=booktitle,
			match=\regexp{.*Advances.*in.*neural.*information.*processing.*},
			replace={Proc. NeurIPS}]
			\step[fieldsource=booktitle,
			match=\regexp{.*Advances.*in.*Neural.*Information.*Processing.*},
			replace={Proc. NeurIPS}]
			\step[fieldsource=booktitle,
			match=\regexp{.*Workshop.*on.* Applications.* of.* Signal.*Processing.*to.*Audio.*and.*Acoustics.*},
			replace={Proc. WASPAA}]
			\step[fieldsource=publisher,
			match=\regexp{.+},
			replace={{}}]
			\step[fieldsource=month,
			match=\regexp{.+},
			replace={{}}]
			\step[fieldsource=location,
			match=\regexp{.+},
			replace={{}}]
			\step[fieldsource=address,
			match=\regexp{.+},
			replace={{}}]
			\step[fieldsource=organization,
			match=\regexp{.+},
			replace={{}}]
		}
	}
}
\begin{document}

\title{\ourmethod{}: Context-aware Token Pruning\\ for Speech Information Retrieval}

\author{%
  \begin{tabular}{@{}c@{\hspace{1em}}c@{\hspace{1em}}c@{\hspace{1em}}c@{}}
    $1^{st}$ Yueqian Lin$^{*}$ & $2^{nd}$ Yuzhe Fu$^{*}$ & $3^{rd}$ Jingyang Zhang & $4^{th}$ Yudong Liu \\[0.1em]
    Duke University & Duke University & Duke University & Duke University \\[0.1em]
    Durham, USA & Durham, USA & Durham, USA & Durham, USA \\[0.1em]
    \texttt{yl768@duke.edu} & \texttt{yf184@duke.edu} & \texttt{jz288@duke.edu} & \texttt{yl817@duke.edu} \\[0.3em]
    $5^{th}$ Jianyi Zhang & $6^{th}$ Jingwei Sun & $7^{th}$ Hai ``Helen'' Li & $8^{th}$ Yiran Chen \\[0.1em]
    Duke University & Duke University & Duke University & Duke University \\[0.1em]
    Durham, USA & Durham, USA & Durham, USA & Durham, USA \\[0.1em]
    \texttt{jz318@duke.edu} & \texttt{js905@duke.edu} & \texttt{hl279@duke.edu} & \texttt{yc278@duke.edu} \\
  \end{tabular}\\[0.3em]
\thanks{
$^{*}$Equal Contribution. This work was supported in part by NSF 2112562 and ARO W911NF-23-2-0224.
}
\\[-8.0ex]
}
\maketitle

\begin{abstract}
While current Speech Large Language Models (Speech LLMs) excel at short-form tasks, they struggle with the computational and representational demands of longer audio clips. 
To advance the model's capabilities with long-form speech, we introduce Speech Information Retrieval (SIR), a long-context task for Speech LLMs, and present SPIRAL, a 1,012-sample benchmark testing models' ability to extract critical details from long spoken inputs. 
To overcome the challenges of processing long speech sequences, we propose \ourmethod{}, a training-free token pruning strategy that uses speech-text similarity and approximated attention scores to efficiently discard irrelevant tokens. 
In SPIRAL, \ourmethod{} achieves accuracy improvements of 29\% and up to 47\% over the original model and the random pruning model at a pruning rate of 20\%, respectively. 
\ourmethod{} can maintain network performance even at a pruning level of 80\%. This highlights the potential of token-level pruning for efficient and scalable long-form speech understanding.

\end{abstract}


\section{Introduction}
Speech Large Language Models (Speech LLMs) represent a significant advancement in speech-language understanding and processing, as they leverage contextual reasoning capabilities of large language models to process audio inputs. Unlike traditional cascaded pipelines, where automatic speech recognition (ASR) and language modeling are handled by separate modules, Speech LLMs unify audio processing, cross-modal fusion, and language modeling in a single architecture~\cite{peng2024survey}. These unified models can perform multiple tasks like speech recognition, speech translation, speaker identification and emotion recognition, while maintaining end-to-end trainability~\cite{chu2024qwen2,DiVA,zhang2023speechgpt,zhan2024anygpt}. 

Despite the broad applications of Speech LLMs, one desirable functionality for these models remains unexplored in existing work.
Specifically, it is the capability of \textit{extracting crucial information within long-context audio}, which we term Speech Information Retrieval (\textbf{SIR}).
SIR is particularly relevant to real-world scenarios, which often require extracting key information from extended audio content, such as meetings, lectures, interviews, and customer service calls.
For instance, the user may want the model (as an AI assistant) to accurately note down the time for a future event mentioned in a long conversation, so as to help them optimize their schedule.
While straightforward to be accomplished by us humans, SIR is non-trivial and challenging for Speech LLMs.
First, the target information will likely exist only in one short audio segment among the whole, extensively long audio inputs.
Precisely recognizing the relevant parts and ignoring the irrelevant parts is intuitively challenging for the models.
Second, as we will discuss later, a more prohibitive limitation for Speech LLMs to perform SIR is their significant computational inefficiency when processing long audio token sequences.

To fill the research gap for SIR, \textit{our first contribution is a concrete task formulation and a rigorously constructed benchmark.}
Note that this effort is necessary and valuable because existing benchmarks for Speech LLMs mostly focus on tasks such as basic speech recognition, translation, and emotion detection, which all emphasize short-term capabilities.
For example, 93\%
of the audio files in the Dynamic-superb phase-2 benchmark~\cite{huang2024dynamic} have a duration of less than 30 seconds.
More recent benchmarks such as MMAU~\cite{sakshi2024mmau} (for complex reasoning) and AudioBench~\cite{wang2024audiobench} (for instruction following) are still limited to short audio inputs (averaging 14.22 and 12.60 seconds respectively).
These benchmarks contain only short audio clips and thus do not reflect the complexity of achieving long-context understanding and extracting precise information from lengthy audio sequences.
To systematically assess the unique challenges posed by SIR, we present \textbf{SPIRAL} (Speech Informational Retrieval and Lookup), a 1,012-sample benchmark specifically crafted to evaluate Speech LLM performance on long-form audio sequences (around 90 seconds in duration). 
On a high level, SPIRAL constructs SIR questions by embedding a critical piece of information within lengthy and potentially distracting dialogues, thereby assessing the model ability to pinpoint and retrieve essential content from long-form inputs.

Preliminary experiments on SPIRAL reveal limitations of current models in handling SIR tasks, due to fundamental architectural constraints. 
Regardless of how audio inputs are encoded, Speech LLMs concatenate the derived audio tokens/embeddings with text tokens for later processing.
However, audio signals typically yield substantially longer token sequences than text inputs, dominating the computational cost and leading to significant inefficiency due to the quadratic complexity of attention with respect to the input length \cite{pmlr-v201-duman-keles23a}.
In fact, most existing models limit the length of input audio files to only 30 seconds \cite{huang2024dynamic} (about 1500 raw tokens when using Whisper~\cite{whisper} for speech encoding, and models typically add adapters to downscale the number of tokens), as otherwise the audio token sequence could easily cause out-of-memory error on GPU.
Obviously, such a limitation is restrictive for Speech LLMs to handle long-form audio inputs longer than 30 seconds.

\begin{figure*}[h]
    \centering
    \includegraphics[width=0.8\textwidth,page=1]{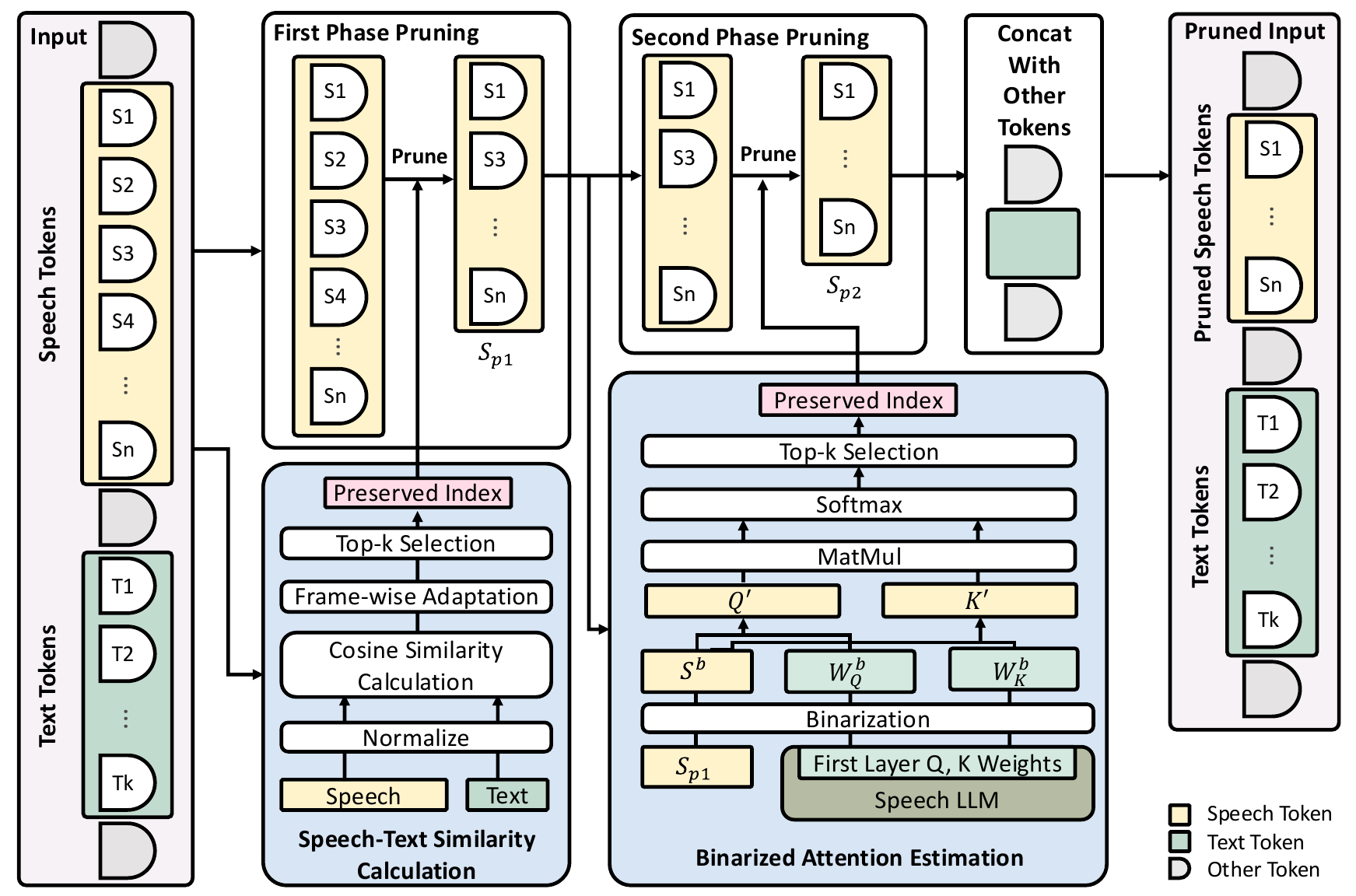}
    \caption{The proposed \ourmethod, with two phases of token pruning.}
    \label{fig:method}
\end{figure*}

To address the limitation, \textit{our second technical contribution is \ourmethod, a training-free token pruning method that enables off-the-shelf Speech LLMs to handle lengthy audio input efficiently and effectively}. 
Unlike existing vision-centric pruning methods (e.g., PruMerge \cite{shang2024llava}) that are incompatible with speech encoders, \ourmethod is specifically designed to preserve the temporal nature of audio signals.
\ourmethod~features a two-phase process, where it first removes semantically irrelevant speech tokens by examining the cosine similarity between speech and text token embeddings, and then further selects the most important tokens by approximating token importance with binarized attention weights from the first layer.
This plug-and-play approach maintains semantic fidelity while substantially reducing computational overhead, making the processing of long audio inputs possible without any additional training upon pre-trained models. 
Our \ourmethod, which, to the best of our knowledge is the first token pruning method for Speech LLMs, achieves nearly 29\% (and 47\%) higher accuracy than the original model (and the random pruning baseline) at a 20\% pruning rate and sustains performance even when pruning 80\% of the input on our SPIRAL benchmark.

\section{Speech Information Retrieval}
\subsection{Task Formulation}

We propose the SIR task to evaluate the ability of Speech LLMs to identify and extract critical information from extended spoken dialogues. 
This task addresses the practical challenge of finding key details within lengthy conversations, akin to finding a ``needle in a haystack,'' which is particularly challenging given most models' constraint of processing only 30-second audio segments.

The task is formulated as follows. Inputs include (1) a long-form speech input $A = {a_1, a_2, \ldots, a_n}$ comprising sequential audio segments $a_i$, where each $a_i$ represents a continuous segment of the spoken dialogue, and (2) a textual query $q$ that targets a specific piece of information mentioned or discussed at some unknown time within the speech. The model must process the entire sequence $A$ to locate the relevant information that answers the query $q$.

This can be formally expressed as
\begin{equation}
    r^* = f(A, q),
\end{equation}
where $r^*$ stands for the correct response, $f$ represents the model's function of processing speech, identifying salient information, and reasoning about the query. The critical information is contained within some segment $a_l$ at position $l$, but this location is not provided to the model explicitly, it must learn to identify and attend to relevant segments while processing the complete sequence.

To ensure accurate evaluation without ambiguity, we structure all queries as multiple-choice questions, following the established practice of multiple existing benchmarks \cite{wang2024audiobench,sakshi2024mmau,huang2024dynamic}.
Note, however, that the proposed SIR task can be easily generalized to open-ended questions as well.
For each query $q$, the model selects from four possible responses $R = \{r_1, r_2, r_3, r_4\}$. This format allows for an objective evaluation of the model's dual capabilities: identifying relevant information in extended audio and understanding its semantic meaning.

\subsection{Benchmark Construction}
We introduce SPIRAL (Speech Information Retrieval And Lookup), a novel benchmark designed to evaluate Speech LLMs' ability to process long and realistic spoken inputs. 
The samples in our dataset feature three representative scenarios, including lectures, meetings, and daily conversations.
Within each scenario, there are various fine-grained and specific topics that ultimately form a diverse and hierarchical topic structure for SPIRAL. 
Unlike existing approaches that simply apply speech synthesis to transform text datasets into speech datasets, we specifically design our data to reflect the unique characteristics of oral communication through a systematic two-stage pipeline, namely transcript generation and speech sample synthesis, in our construction.

\noindent\textbf{Transcript Generation}
The transcript generation process employs the advanced capabilities of GPT-4o to simulate dialogues that are indistinguishable from natural human conversations. This simulation covers a wide array of topics ranging from everyday life scenarios to professional exchanges and social interactions. The methodology unfolds as follows:
\begin{enumerate}
    \item \textbf{Topic Curation:} A comprehensive array of topics is meticulously selected to capture the breadth and complexity of human interactions, with hierarchial orgnization to ensure diverse coverage across domains.
    \item \textbf{Dialogue Generation:} Using GPT-4o, we generate multi-turn dialogues incorporating natural speech elements (fillers like ``uh'' and ``oh'') to enhance authenticity. Our prompt engineering specifically guides the model to create realistic conversational dynamics with variable turn lengths and contextual continuity. Multiple-choice questions are generated for evaluation purposes.
\end{enumerate}
\noindent\textbf{Speech Sample Synthesis}
The speech synthesis process utilizes the capabilities of StyleTTS 2 \cite{li2024styletts}, a state-of-the-art zero-shot text-to-speech engine trained on the LibriTTS dataset \cite{zen2019libritts}. Our synthesis pipeline comprises the following steps:
\begin{enumerate}
    \item \textbf{Speaker Selection:} Speakers are randomly selected from the train-clean-100 dataset in LibriTTS with balanced gender representation from the LibriTTS dataset to ensure diversity and avoid gender bias in our audio samples.
    \item \textbf{Speech Generation:}  Using StyleTTS 2, we generate speech with fixed diffusion steps and embedding scale parameters. Dialogue turns are concatenated to create continuous speech while preserving conversational flow. 
\end{enumerate}

The SPIRAL dataset is open-source,\footnote{The dataset and its construction details can be accessed through our \href{https://speechprune.github.io/}{website}.} facilitating further research on SIR tasks. In addition, we propose SPIRAL-H, a challenging subset consisting of 401 cases in which the original Qwen-2 Audio model used in our experiments fails completely, achieving 0\% accuracy.

\subsection{Quality Assessment}

The generated SPIRAL dataset contains 1,012 samples, with an average duration of 87.89 seconds.
To assess the quality of our generated SPIRAL dataset, we evaluate the synthesized samples using two complementary metrics: automatic speech recognition accuracy via Whisper-v3-large \cite{whisper}, which achieves a word error rate of 0.0389, and perceptual quality via UTMOS-22~\cite{saeki22c_interspeech}, a widely used surrogate objective metric of mean opinion score (MOS), yielding a predicted MOS of 3.91 in a five-point-scale. These metrics respectively quantify the transcription accuracy of the speech content by a state-of-the-art recognition system and the naturalness/human-likeness of the speech, as evaluated by a perceptual quality model. SPIRAL demonstrates strong performance in both metrics.

\section{SpeechPrune}
\subsection{Preliminaries}
\noindent\textbf{Audio Encoder} Speech LLMs typically consist of an audio encoder (such as Whisper \cite{whisper}) which transforms raw audio with high sampling rates into lower-dimensional embeddings.
Taking Whisper as an example, an audio input (with maximum length) is first processed and transformed into an 80-channel melspectrogram in the time-frequency domain. This 80-channel melspectrogram, generated with a window size of 25 ms and a hop size of 10 ms, is then fed into the Transformer-based encoder. A pooling layer with a stride of two follows to reduce the length of the audio representation.
As a result, each frame of the encoder output approximately corresponds to a 40ms segment of the original audio signal.
Thus, a 30-second audio yields 750 encoding embeddings.
This temporal correspondence between audio frames and encoder outputs provides a natural foundation for our frame-level pruning strategy, as we can leverage the inherent structure of how speech information is encoded to maintain temporal coherence during pruning.

\noindent\textbf{Language Modeling}
After extracting the audio token, it is typically projected by an MLP \cite{gong2023listen} or Q-Former \cite{tang2023salmonn} to align the feature-wise dimensionality with text tokens.
The audio token is then concatenated with the text token and other system prompts before being input to the LLM backbone \cite{das2024speechverse}.
In transformer-based models, the self-attention mechanism for each layer is computed as
\begin{equation}
\text{Attention}(\mathbf{Q}, \mathbf{K}, \mathbf{V}) = \text{softmax}\left(\frac{\mathbf{Q}\mathbf{K}^T}{\sqrt{d_k}}\right)\mathbf{V},
\end{equation}
where $\mathbf{Q}$, $\mathbf{K}$, and $\mathbf{V}$ are the query, key, and value  derived from the input sequence $\mathbf{X}$ through learnable projections:

\begin{equation}
\mathbf{Q} = \mathbf{X}\mathbf{W}_Q, \quad \mathbf{K} = \mathbf{X}\mathbf{W}_K, \quad \mathbf{V} = \mathbf{X}\mathbf{W}_V.
\end{equation}

The quadratic complexity $O(n^2)$ of self-attention mechanisms \cite{vaswani2017attention, pmlr-v162-hua22a} makes the length of audio tokens a critical computational bottleneck.
For instance, a 10-minute conversation with approximately 15,000 tokens requires 58.66 TFLOPS for Qwen-2 network \cite{chu2024qwen2}, highlighting the need for efficient pruning strategies \cite{kim2022learned,shang2024llava}.

\subsection{\ourmethod{} Methodology}
We propose a two-phase token pruning approach, as shown in Fig.~\ref{fig:method} and the following parts.

\noindent\textbf{First Phase Pruning by Token-Text Similarity} The first phase utilizes the correlation between audio and text tokens to identify semantically important audio segments.
Recent research has shown that such audio-text token alignment enables effective cross-modal reasoning in speech-language models \cite{DiVA}.
More formally, we process the input to get speech embedding $\mathbf{S}\in\mathbb{R}^{N\times D}$ and text embedding $\mathbf{T}\in\mathbb{R}^{L\times D}$, where $N$ is the number of speech tokens before pruning, $L$ is the number of text tokens, and $D$ is the embedding dimensionality. 
Here, we only consider real text query as $\mathbf{T}$ and exclude system prompt and special tokens.
The token-level similarity matrix $\mathbf{F} \in \mathbb{R}^{N \times L}$ between speech and text tokens is computed using cosine similarity:

\begin{equation}
\mathbf{F} = \frac{\mathbf{S}}{\|\mathbf{S}\|_2} \cdot \frac{\mathbf{T}^{\top}}{\|\mathbf{T}\|_2}.
\end{equation}

We introduce an adaptive frame-level approach to enhance natural continuity and temporal correspondence. This method evaluates speech segments as one-second frames, aligning with the delta-band oscillations (1-2 Hz) that naturally process lexical and phrasal units in speech perception \cite{giraud2012cortical}. Given the speech embedding $\mathbf{S}$, we obtain $m = \lceil N/f \rceil$ frames, where $f$ is the frame size per second. For each frame $i$, the mean similarity score across text tokens is first computed, followed by frame-wise accumulation:
\begin{equation}
\hat{\mathbf{F}}_i = \sum{j=0}^{f-1} \operatorname{mean}(\mathbf{F}_{i\cdot f+j,:}, \text{axis}=1),
\end{equation}
where $\mathbf{F}_{i\cdot f+j,:}$ represents the similarity scores between the $j$-th token in frame $i$ and all text tokens. Token retention within each frame is determined by a softmax function applied to the accumulated frame scores:
\begin{equation}
\mathbf{p} = \operatorname{softmax}(\hat{\mathbf{F}}).
\end{equation}
The expected number of tokens to retain from each frame is
\begin{equation}
n_i = \left\lfloor N p_i \right\rfloor,
\end{equation}
where $N$ denotes the overall number of tokens to be retained. 
For each frame $i$, we select the top-$n_i$ tokens based on their mean similarity scores:
\begin{equation}
\begin{array}{r}
\text{indices}_{\text{first},i} = \operatorname{topk}(\operatorname{mean}(\mathbf{F}_{i\cdot f:(i+1)\cdot f,:}, \text{axis}=1), n_i), \\ \text{for } i = 1,\ldots,m,
\end{array}
\end{equation}
where $\mathbf{F}_{i\cdot f:(i+1)\cdot f,:}$ represents the similarity scores of tokens within frame $i$ across all text tokens. 

The speech token remaining after first phase pruning is:

\begin{equation}
\mathbf{S}_{p1} = \mathbf{S}[\cup_{i=1}^m \text{indices}_{\text{first},i}].
\end{equation}

\noindent\textbf{Second Phase Pruning by Binarized Attention Estimation}
Building on the first-phase pruning results, we introduce a second pruning phase to further select important tokens based on approximated attention scores. This phase exclusively focuses on speech tokens, as the text-speech relationships have already been captured in the first pruning phase, enabling efficient modeling of internal dependencies within speech segments while minimizing computational overhead. The second phase utilizes the binarized attention from the network's first transformer layer. Specifically, we compute the scores using the signed binarized Query and Key weights, and also the pruned speech embeddings:

\begin{equation}
(\mathbf{W}_Q^b\text{, } \mathbf{W}_K^b\text{, } \mathbf{S}^b) = \operatorname{sign}(\mathbf{W}_Q\text{, }\mathbf{W}_K\text{, }\mathbf{S}_{p1}).
\end{equation}
Then the approximate attention scores are computed through binarized matrix operations:
\begin{equation}
\mathbf{Q}' = \mathbf{S}^b \mathbf{W}_Q^b\text{, }\mathbf{K}' = \mathbf{S}^b \mathbf{W}_K^b,
\end{equation}
\begin{equation}
\mathbf{A} = \operatorname{softmax}(\frac{\mathbf{Q}' {\mathbf{K}'}^\top}{\sqrt{d_k}}).
\end{equation}
The final token selection is determined by
\begin{equation}
\mathbf{S}_{p2} = \mathbf{S}_{p1}[\operatorname{topk}(\operatorname{mean}(\mathbf{A}, \text{axis}=1), k)].
\end{equation}

This simplified attention mechanism accounts for less than 1\% of the network's total computational complexity, which is highly efficient. The final pruned input merges selected audio tokens $\mathbf{S}_{p2}$ with other essential tokens.
\section{Experiments}
We conduct our main experiments using Qwen-2 Audio \cite{chu2024qwen2}, a state-of-the-art Speech LLM with extensive speech understanding task coverage. 
Our primary results are presented in Section~\ref{sec:main_experiments}, with  qualitative analyses discussed in Section~\ref{sec:qualitative_analysis}. 
Additionally, we perform ablation studies examining the performance impact of each pruning phase in Section~\ref{sec:ablation_studies}. Finally, Section~\ref{sec:generalization_analysis} demonstrates the generalizability of our proposed method across different models and benchmarks.
\subsection{Main Experiments}
\label{sec:main_experiments}
\begin{table}
\centering
\small
\caption{Comparison of different audio pruning methods across various metrics. PR: Pruning Rate, TF: TFLOPS, PT: Prefill time (ms), TM: Total memory (GB), SA: Storing activation (GB), RAP: Random Audio Pruning, RAC: Random Audio Cropping.}
\label{tab:main_results}
\scalebox{0.75}{
\begin{tabular}{lcccccccc}
\toprule[2pt]
Method & PR & TF $\downarrow$ & PT $\downarrow$ & TM $\downarrow$ & SA $\downarrow$ & SPIRAL $\uparrow$ & SPIRAL-H $\uparrow$ \\
\midrule[1pt]
Original & -- & 12.2 & 779 & 13.40 & 0.19 & 60.38\% & 0\% \\
\midrule[0.5pt]
RAP & \multirow{3}{*}{0.2} & \multirow{3}{*}{10.06} & \multirow{3}{*}{662} & \multirow{3}{*}{13.32} & \multirow{3}{*}{0.15} & 42.49\% & 21.45\% \\
RAC &  & & & & & 65.71\% & 48.13\% \\
\textbf{Ours} &  & & & & & \textbf{89.23\%} & \textbf{81.64\%} \\
\midrule[0.5pt]
RAP & \multirow{3}{*}{0.4} & \multirow{3}{*}{7.93} & \multirow{3}{*}{511} & \multirow{3}{*}{13.24} & \multirow{3}{*}{0.11} & 42.89\% & 22.19\% \\
RAC &  & & & & & 62.45\% & 41.90\% \\
\textbf{Ours} &  & & & & & \textbf{85.97\%} & \textbf{76.43\%} \\
\midrule[0.5pt]
RAP & \multirow{3}{*}{0.6} & \multirow{3}{*}{5.79} & \multirow{3}{*}{419} & \multirow{3}{*}{13.17} & \multirow{3}{*}{0.07} & 42.39\% & 21.45\% \\
RAC &  & & & & & 58.20\% & 35.41\% \\
\textbf{Ours} &  & & & & & \textbf{75.89\%} & \textbf{63.77\%} \\
\midrule[0.5pt]
RAP & \multirow{3}{*}{0.8} & \multirow{3}{*}{3.66} & \multirow{3}{*}{278} & \multirow{3}{*}{13.09} & \multirow{3}{*}{0.04} & 45.26\% & 23.19\% \\
RAC &  & & & & & 55.83\% & 33.67\% \\
\textbf{Ours} &  & & & & & \textbf{62.45\%} & \textbf{46.15\%} \\
\bottomrule[2pt]
\end{tabular}
}
\end{table}

\noindent\textbf{Setup}
We evaluate our method using Qwen-2 Audio, comparing our \ourmethod\ method against several baselines, comparing our two-phase pruning strategy (\ourmethod) against three baselines: (1) Original: full audio trimmed at 30 seconds (750 tokens); (2) RAP: random audio pruning that selects non-contiguous segments to reach target rate; and (3) RAC: random audio cropping that selects a single contiguous segment at target rate. 
Our \ourmethod's two-phase pruning strategy is set as follows: the first phase prunes the input tokens to match the original method's input length (which is 750 tokens), while the second phase removes additional tokens according to the specified pruning rate.
We evaluate computational efficiency using TFLOPS\footnote{Calculated using calflops: \url{https://github.com/MrYxJ/calculate-flops.pytorch}}, measure prefill time on a Quadro RTX6000 GPU, and assess memory usage (total and activations) using LLM-Viewer~\cite{yuan2024llm}.

\noindent\textbf{Results}
Our results in Table \ref{tab:main_results} demonstrate that \ourmethod\ outperforms all baseline methods across different pruning rates, achieving $89.23\%$ accuracy on the SPIRAL benchmark and $81.64\%$ on the more challenging SPIRAL-H subset when pruning $20\%$ of the input, compared to the original model's $60.38\%$ and $0\%$ respectively. SPIRAL-H is particularly notable as it consists of 401 challenging cases where the original model completely fails (having $0\%$ accuracy). Even with aggressive pruning ($80\%$ pruning rate), our method maintains the network accuracy while reducing $70\%$ computational costs (from $12.2$ to $3.66$ TFLOPS), $64\%$ prefill time (from $779$ to $278$ ms), and saving $79\%$ activation storage (from $0.19$ to $0.04$ GB) compared to the original model. The inherent randomness of RAP and RAC often fails to identify crucial information, resulting in an inconsistent relationship between network accuracy and pruning rate. In contrast, \ourmethod\ demonstrates a more systematic approach by effectively selecting critical information, which leads to a more predictable and gradual decline in network performance as pruning rates increase. Furthermore, \ourmethod\ consistently outperforms random pruning strategies in terms of accuracy, with up to $46.74\%$ in SPIRAL and $60.19\%$ in SPIRAL-H.
\vspace{-0.5em}

\subsection{Qualitative Analysis}
\label{sec:qualitative_analysis}

\begin{figure}[t]
   \centering
   \includegraphics[width=1\linewidth]{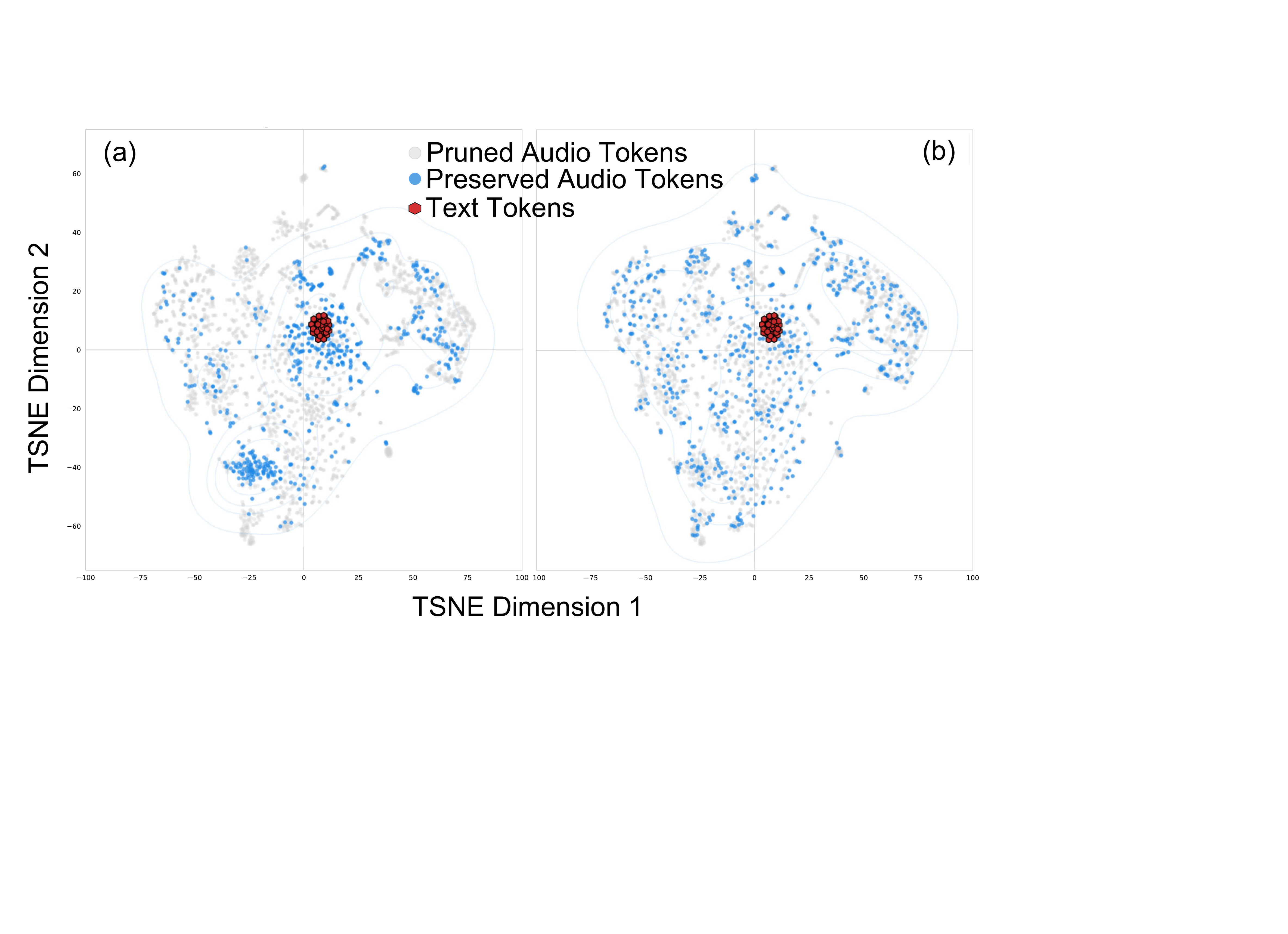}
   \caption{Qualitative analysis of token embeddings via t-SNE visualization, where high-dimensional embeddings are projected into 2D space for visualization. (a) \ourmethod{} (b) Random pruning. Gray, blue, and red points represent pruned audio tokens, preserved audio tokens, and text tokens, respectively.}
   \label{fig:qualitative_analysis}
\end{figure}

To visualize the effectiveness of our pruning strategy, we project token embeddings from one sample in the SPIRAL dataset into a 2D space using t-SNE visualization, comparing distributions between \ourmethod\ and RAP (Fig. \ref{fig:qualitative_analysis}). Our method demonstrates more structured token selection, where preserved audio tokens (blue) exhibit stronger clustering around text tokens (red) compared to the scattered distribution in random pruning, suggesting effective retention of semantically relevant audio information. This visualization corroborates our quantitative results, showing \ourmethod's capability to maintain semantic relationships in the pruned representation.

\subsection{Ablation Studies}
\label{sec:ablation_studies}

\begin{figure}[t]
    \centering
    \includegraphics[width=0.9\linewidth]{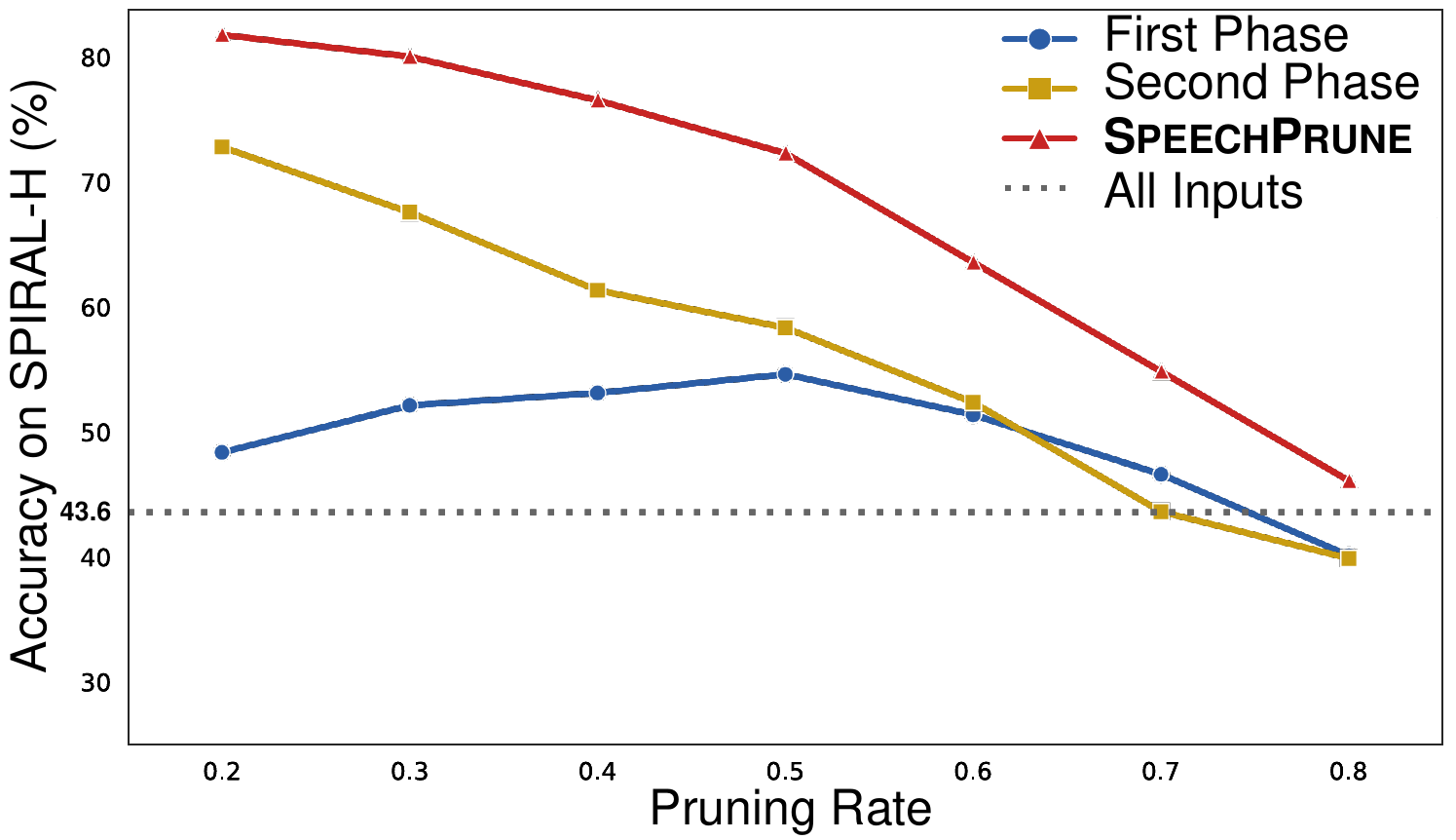}
    \caption{Ablation study comparing different pruning strategies on SPIRAL-H dataset.
    The plot shows the accuracy of three approaches: first phase only,
    second phase only, and \ourmethod.
    The dotted line shows the accuracy when using the complete, unpruned set of input tokens.}
    \label{fig:ablation_study}
\end{figure}

To evaluate the effectiveness of our two-phase pruning approach, we conduct ablation studies on the SPIRAL-H dataset. We examine three variants of our method: using only the first phase pruning, using only the second phase pruning, and the complete two-phase approach.  
Fig.~\ref{fig:ablation_study} presents the performance comparison across different pruning rates. When using the complete set of unpruned input tokens, the model achieves an accuracy of 43.6\%. The combined approach consistently outperforms both individual pruning phases across most pruning rates, achieving peak performance of $81.64\%$ at $0.2$ pruning rate compared to $48.13\%$ and $72.45\%$ for first phase and second phase only, respectively. This significant improvement over the original model's $0\%$ accuracy on SPIRAL-H indicates that our pruning strategy not only reduces computational cost but also enhances the model's ability to identify and process critical information. Second, we observe interesting behavioral patterns for each variant: the first phase only approach shows relatively stable but lower performance ($45$-$55\%$), while the second phase only method starts with higher accuracy but degrades more rapidly as pruning rate increases. Finally, the combined approach exhibits the most robust performance, maintaining superior accuracy until around $0.7$ pruning rate, after which all methods converge to similar performance levels. This suggests that our two-phase design leverages complementary information from both token-level similarity and attention patterns, resulting in more robust and efficient pruning even on challenging cases where the original model fails.

\subsection{Generalization Analysis}
\label{sec:generalization_analysis}
To evaluate the generalization capability of our method, we test \ourmethod\ on both different benchmarks and a different Speech LLM model. For additional benchmarks, we select two representative long-form speech understanding datasets: DREAM-TTS and CN-College-Listen. DREAM-TTS is derived from the text-based dialogue comprehension dataset DREAM~\cite{DREAM}, converted to speech using state-of-the-art TTS technology with 60 different speakers while maintaining gender consistency as described by \cite{wang2024audiobench}. CN-College-Listen is sourced from WavLLM's~\cite{hu2024wavllm} test set, comprising English listening comprehension questions from China's national college entrance examinations. For both datasets, we specifically use test samples that exceed 60 seconds in length to evaluate long-form speech understanding capabilities.

We also evaluate our method on DiVA~\cite{DiVA}, a recently proposed Speech LLM trained without instruction data using text-only LLM responses as self-supervision. As shown in Table~\ref{tab:benchmark_comparison}, \ourmethod\ demonstrates consistent improvements across all benchmarks and models. For Qwen-2 Audio with $0.2$ pruning rate, our method improves accuracy from $53.69\%$ to $65.19\%$ on DREAM-TTS and from $52.91\%$ to $62.86\%$ on CN-College-Listen. When applied to DiVA with $0.15$ pruning rate, \ourmethod\ similarly enhances performance across all three benchmarks, demonstrating its effectiveness even on models trained with different paradigms. These results suggest that our pruning strategy generalizes well across different types of speech understanding tasks and model architectures, even though these benchmarks were not originally designed specifically for SIR tasks.
\begin{table}[h]
\centering
\small
\caption{Performance comparison on SPIRAL, DREAM-TTS (DTTS), and CN-College-Listen (CCL) benchmark using Qwen-2 Audio (pruning rate: 0.2) and DiVA (pruning rate: 0.15). The symbol * indicates results obtained on a subset of the benchmark where the audio duration exceeds 60 seconds.}
\scalebox{0.85}{
\begin{tabular}{lccc}
\toprule
\multirow{2}{*}{Model} & \multicolumn{3}{c}{Accuracy (\%)} \\
 & SPIRAL & DTTS* & CCL* \\
\midrule
Qwen-2 Audio & 60.38 & 53.69 & 52.91 \\
\rowcolor{rowcolor} \quad + \ourmethod & \textbf{89.23} & \textbf{65.19} & \textbf{62.86}\\
DiVA & 48.62 & 45.72 & 55.24\\
\rowcolor{rowcolor} \quad + \ourmethod & \textbf{57.51} & \textbf{53.10} & \textbf{56.19}\\
\bottomrule
\end{tabular}}
\label{tab:benchmark_comparison}
\end{table}

\section{Conclusion}
In this work, we introduced the SIR task to target long-form speech comprehension, presented SPIRAL as a benchmark for evaluating such capabilities, and proposed \ourmethod
, a training-free token pruning method leveraging speech-text similarity and approximate attention. Experimental results showed that \ourmethod\ not only reduces computational costs but can also enhance model performance, achieving network accuracy improvements of nearly $29\%$ and up to $47\%$ over the original model and the random pruning model, respectively. While promising, further exploration is needed to improve robustness under diverse audio conditions, explore additional token selection methods, and adapt pruning strategies to specific input characteristics or fine-tuned models.
\section*{References}
{
\printbibliography
}
\end{document}